\documentclass[twocolumn]{article}
\usepackage{abstract}
\usepackage{amsfonts}
\usepackage{amssymb}
\newcommand{\sep}{\text{sep}}
\renewcommand{\H}{{\bf H}}
\renewcommand{\v}{{\bf v}}
\newcommand{\1}{{\bf 1}}
\newcommand{\0}{{\bf 0}}

\newcommand{\x}{{\bf x}}
\newcommand{\text}{\mbox}
\newcommand{\tr}{\text{Tr}}

\renewcommand{\sep}{\text{sep}}
\newtheorem{Lemma}{Lemma}
\newtheorem{Theorem}{Theorem}
\newtheorem{Cor}{Corollary}
\newcommand{\beq}{\begin{equation}}
\newcommand{\eeq}{\end{equation}}
\newcommand{\beqa}{\begin{eqnarray}}
\newcommand{\eeqa}{\end{eqnarray}}

\def\union{\cup}
\def\intersect{\cap}

\def\QED{$\Box$}

\title{\Large \bf Cloning and Broadcasting in Generic Probabilistic Models}
\author{Howard Barnum$^1$ \and Jonathan Barrett$^2$ \and Matthew Leifer$^{2,3}$ \and Alexander Wilce$^4$\\\\
$^1$\small\em CCS-3: Modeling, Algorithms, and Informatics, Mail Stop B256,
\\ \small \em Los Alamos National Laboratory, Los Alamos, NM 87545 USA.\\\\
$^2$\small\em Perimeter Institute for Theoretical Physics, 31 Caroline Street North, Waterloo, Ontario,\\\small\em Canada, N2L 2Y5\\\\
$^3$\small\em Centre for Quantum Computation, Department of Applied Maths and Theoretical Physics,\\\small\em University of Cambridge, Wilberforce Road, Cambridge, CB3 0WA, UK\\\\
$^4$\small\em Department of Mathematical Sciences, Susquehanna University, Selinsgrove, PA 17870 USA.\\
}

\date{\today}

\begin{document}

\twocolumn[

\maketitle

\begin{onecolabstract} We prove generic versions of the no-cloning
and no-broadcasting theorems, applicable to essentially {\em any}
non-classical finite-dimensional probabilistic model that satisfies a
no-signaling criterion.  This includes quantum theory as well as
models supporting ``super-quantum'' correlations that violate the Bell
inequalities to a larger extent than quantum theory.  The proof of our
no-broadcasting theorem is significantly more natural and more
self-contained than others we have seen: we show that a set of states
is broadcastable if, and only if, it is contained in a simplex whose vertices are
cloneable, and therefore distinguishable by a single measurement.  This
necessary and sufficient condition generalizes the quantum requirement that
a broadcastable set of states commute.  \\

\end{onecolabstract}]

\section{Introduction}

% *ML20/11/06* Rewrote Howard's new introductory paragraph
The growth of quantum information science has led many to wonder
which aspects of quantum mechanics are responsible for its enhanced
information processing powers.  Some have compared quantum and
classical theories in frameworks broad enough to encompass both of
them and more \cite{Hardy2001a, Hardy2001b, Barnum2002a,
Barnum2003a, Barnum2004a, Barrett:2005, d'Ariano2006a}, and others
have constructed toy theories that capture qualitative features of
quantum information protocols \cite{Hardy99a,Spekkens:2004,
Smolin2003a}. Beyond simply understanding the conceptual sources of
the power of quantum theory, researchers have become interested in
information-processing as a source of axioms that could characterize
probabilistic physical %AW: added the phrase "probabilistic physica:
theories \cite{Fuchs:2002, Fuchs:2003,
Clifton:2003, Barnum2002a, Barnum2003a, Spekkens:2004,
Barrett:2005}, shedding light on the conceptual essence of quantum
mechanics and potentially giving new stimulus to the longstanding
program \cite{Mackey:1963, Ludwig64a, Ludwig67a, Ludwig81a,
Ludwig:1983, Araki80a} of axiomatic characterization of quantum
theory.  It has even been suggested that this approach might ease
the integration of quantum theory with general relativity and
gravitation \cite{Hardy:2005}

As part of this development several authors, notably Barrett
\cite{Barrett:2005} and Spekkens \cite{Spekkens:2004}, have recently
taken up the question of how far the information-theoretic novelties
presented by quantum mechanics are in fact generic in other types of
probabilistic theory.
%including those in which states
%are manifestly states of knowledge.
Spekkens constructs an ingenious ``toy model'' in which a limitation
on the amount of knowledge available to observers is sufficient to
yield, among many other things, a no-cloning property. Working in a
framework in which essentially any finite-dimensional compact convex
set counts as a state space, Barrett shows that universal
probabilistic cloning is impossible in {\em any} non-classical
finite-dimensional probabilistic theory.

A major motivation for Barrett's work was to come up with a
reasonable physical framework in which arbitrary nonsignaling
correlations may be obtained from measurements on a bipartite
system.  Such correlations can be more non-local than quantum theory
allows, and include the super-quantum correlations that have come to
be known as Popescu-Rohrlich (PR) boxes, or Non-Local Machines
\cite{Khalfi:1985, Popescu:1994, Barrett:2005a, Wolf:2005,
Short:2005, Buhrman:2005, Broadbent:2005, Barrett:2005b, Jones:2005,
Dam:2000, Dam:2005, Brassard:2005}.  His framework is based on that
of Hardy, and in developing it Barrett and Hardy have essentially
reinvented the finite-dimensional version of a much older framework
for generalized probabilistic models, based on convex sets
\cite{Mackey:1963, Ludwig64a, Ludwig67a, Ludwig81a, Ludwig:1983,
Ludwig:1985, Foulis:1981, Beltrametti97a, Gudder99b, Holevo:1983},
which grew out of attempts to axiomatize quantum theory within the
quantum logic tradition, and which we adopt here.

Popescu and Rohrlich \cite{Popescu:1994} originally raised the
question of why nature does not allow super-quantum correlations,
given that they would not violate relativistic causality.  In this
regard, it is important to distinguish the unique features of
quantum theory from those that would still hold in theories
permitting more general correlations.  Placing PR boxes within a
framework that also includes quantum theory and classical
probability theory as special
cases,%HB added and classical probability AW: added theory
helps to understand the common features that these have been found
to exhibit (see \cite{Barrett:2005} for a discussion of many of
these).

In this paper, we completely characterize the sets of states that
can be cloned or broadcast in any finite-dimensional probabilistic
model within the convex sets framework, obtaining along the way a
simple, natural, and self-contained proof of the quantum
no-broadcasting theorem that is substantially simpler than the
original proof of Barnum, Caves, Fuchs, Jozsa, and Schumacher
\cite{Barnum:1996}, and substantially more intuitive and
self-contained than that based on Lindblad's Theorem
\cite{Lindblad:1999} (which, however, provided some suggestive
ideas).

In section 2, we sketch the standard framework for generalized
probability theory, in which arbitrary compact convex sets are
construed as state-spaces. We restrict our attention, in the main,
to finite-dimensional state spaces. In this context, a state space
is classical iff it is a simplex. In section 3, we discuss the
maximal, or injective, tensor product of convex sets, pointing out
along the way some familiar aspects of entanglement (e.g.,
entanglement monogamy) that hold generically for all non-classical
models. In section 4, we prove our generic no-cloning theorem. We
show that the set of states cloned by an affine mapping must be
distinguishable from one another, with certainty, by a single
observable. It follows that only when the state-space is a simplex
is it possible to clone all pure states.

%[!!!!****!!!!WARNING!!!****!!!!  This next paragraph has to be written in light of Howard's new contribution.
%I'm leaving this until that section is sorted out. AW: it seems alright -- but perhaps we should refer to the appendix?]

In section 5, we show that the set of states broadcast by an affine
mapping is contained in a possibly larger set of states, the extreme
points of which are cloned by an affine map. It follows that the
extreme points of this larger set are distinguishable.  In fact we
show that a set of states is broadcastable if, and only if, it is
contained in a simplex whose vertices are jointly distinguishable.
In the quantum-mechanical setting, convex combinations of
distinguishable states commute, so we obtain the quantum
no-broadcasting theorem as a corollary.  Finally, we extend this
result to show that for {\em any} affine map, the set of states it
broadcasts is a (possibly empty) simplex whose vertices are
distinguishable states. To prove this, we use an extension of the
classical Perron-Frobenius theory for (possibly reducible)
non-negative real square matrices. The necessary technical apparatus
is collected in an appendix.

%These results should be taken as part of a general programme,
%implicit not only in recent work such as [B, Sp], but also in much
%of the existing ``quantum-logical" literature, of sharply
%distinguishing between probabilistic phenomena that are specifically
%quantum, and those that are merely non-classical. As our results
%show, no-cloning and no-broadcasting are of the latter sort.
%This should be contrasted with the view expressed, e.g., in [ref]:
%\begin{quote}
%... the most distinctive feature of quantum information, which makes
%it a "thing" rather than ``knowledge", is that it cannot be shared.
%\end{quote}
%On the contrary, it is trivial to construct ``toy models" in the
%sense of Spekkens, for which states are manifestly {\em just} states of
%knowledge; nevertheless, in theories based on such models, knowledge
%is not always accessible. Our results confirm what one would expect:
%limitations on each party's access to information place collateral limitations
%on the ability of two parties to share information. \\
\newpage

\section{The Framework}

To survey all possible probabilistic theories requires some
altitude. That is, one needs to work in a mathematical framework
that imposes only the most minimal constraints on the structure of
probabilistic models. Such a framework was constructed, for just
this purpose, by Mackey \cite{Mackey:1963} in the late 1950s;
refinements and stylistic variants of this can be found in the work
of many other authors, including Ludwig \cite{Ludwig64a, Ludwig67a,
Ludwig81a, Ludwig:1983, Ludwig:1985}, Foulis and Randall
\cite{Foulis:1981}, Beltrametti and Bugajski \cite{Beltrametti97a},
Gudder et. al.
\cite{Gudder99b}, and Holevo \cite{Holevo:1983}. %AW: I'd like also to cite Davies and Lewis here.. save for v2.
The framework developed by Hardy \cite{Hardy2001a,Hardy2001b} (see also \cite{Mana2004a}) for an axiomatic derivation of quantum mechanics is essentially a finite-dimensional version.
What follows is simply a sketch of this common, more or less canonical, framework. \\

\noindent {\em States and Effects}\\

We assume that a physical system is characterized by its state-space
$\Omega$, which we take to be convex. We write $A(\Omega)$ for the
space of all affine linear functionals $f : \Omega \rightarrow {\Bbb
R}$, and $A(\Omega)_+$ for the space of all nonnegative linear
functionals $f : \Omega \rightarrow {\Bbb R}_+$. Note that
$A(\Omega)$ is an ordered linear space, with $f \leq g$ iff
$f(\omega) \leq g(\omega)$ for all $\omega \in \Omega$. The {\em
order unit} of $A(\Omega)$ is the unit functional $u$ given by
$u(\omega) = 1$ for all $\omega \in \Omega$; the {\em unit interval}
in $A(\Omega)$ is the set $[0,u]$ consisting of all functionals $a
\in A(\Omega)$ satisfying $0 \leq a \leq u$ (in the pointwise
ordering on $\Omega$).
%hb 1 --> u, assuming \leg is the ordering of the space of functionals;
%if we keep \leq 1 then we should evaluate it on \omega and say for all \omega
%AW: added parenthetical phrase above to clarify the ordering.

We interpret each $a \in [0,u]$ as representing an ``effect'' --
that is, some possible event or occurrence associated with the
system -- and $a(\omega)$, as the {\em probability} of this
occurrence when the system is in state $\omega$. There is a natural
embedding of $\Omega$ in $A(\Omega)^{\ast}$, given by $\omega
\mapsto \hat{\omega}$, where $\hat{\omega}(a) = a(\omega)$ for all
$a \in A(\Omega)$. Henceforth, we identify $\omega$ with
$\hat{\omega}$, writing $\omega(a)$ in place of $a(\omega)$, as this
is in better keeping with the the idea of states assigning
probabilities to effects (rather than effects assigning expected
values to states).

We write $V(\Omega)$ for the span of $\Omega$ in $A(\Omega)^\ast$.
The space $V$ is ordered by the cone $V_{+}$ consisting of of all
$\mu \in V$ with $\mu(a) \geq 0$ for every $a \in A(\Omega)_{+}$.
Equivalently, $\mu \in V_{+}$ iff $\mu$ is a non-negative multiple
of a state $\omega \in \Omega$. Accordingly, we call elements of
$V(\Omega)$ {\em weights}. We say that $\Omega$ is
finite-dimensional iff $V(\Omega)$ is finite-dimensional, and {\em
compact} iff $\Omega$ is compact in the weakest topology making
% *ML20/11/06* Removed parentheses around (evaluation at).
evaluation at each $a \in [0,u]$ continuous.  For the remainder of
this paper, we make the standing assumption that all state spaces
are finite-dimensional and compact (equivalently, closed)
as subsets of $A(\Omega)^{\ast}$.
This guarantees that $\Omega$ is the closed convex hull of its extreme points, which are referred to as \emph{pure} states.  \\
%AW: Rewrote the preceding sentence slightly

\noindent{\em Examples}\\

In constructing examples, one often begins with a {\em test space}
(or {\em manual}) \cite{Foulis:1981, Klay:1988, Klay:1987}: that is,
a collection $\frak A$ of (not necessarily disjoint) sets $E, F,
....$, called {\em tests}, interpreted as the outcome-sets of
various measurements. Let $X = \bigcup {\frak A}$ be the set of all
outcomes of all tests $E \in {\frak A}$. A {\em state} on $\frak A$
is defined to be a mapping $\omega : X \rightarrow [0,1]$ summing
independently to $1$ over each $E \in {\frak A}$. The collection
$\Omega({\frak A})$ of all such states is obviously convex. If each
$E \in {\frak A}$ is finite, then it is also compact in the topology
of pointwise convergence on $X$ \cite{Wilce:1992}. %AW: Not certain this is the right reference ..
A state is {\em deterministic} ({\em dispersion-free}) iff its value
on each outcome $x \in E$ is either $0$ or $1$.

(a) If ${\frak A}$ consists of a single test $E$, with a finite number
of outcomes then $\Omega({\frak A})$ is the set of all classical
probability distributions over $E$.  This is a simplex, which we
denote by $\Delta(E)$.

(b) If ${\frak A}$ consists of two two-outcome tests $E_0 =
\{a_{00},a_{01}\}$ and $E_1 = \{a_{10},a_{11}\}$, then $\Omega({\frak
A})$ is a square.  The index $i$ in $a_{ij}$ can be thought of as the
``input'', corresponding to the choice of measurement to be performed
on the system, and the index $j$ can be thought of as a binary
``output''.  Then, the states $\omega \in \Omega({\frak A})$ can be
thought of as conditional probability distributions (or equivalently
$2 \times 2$ stochastic matrices) where $p(\mbox{output} = j|
\mbox{input} = i) = \omega(a_{ij})$, and any conditional probability
distribution likewise defines a valid state. The four vertices of the
state space are the four deterministic states corresponding to the
choice of a definite output for each possible input.  Clearly, this
construction can be repeated for a test space with any number of
nonoverlapping tests, the resulting state space being an appropriate
set of conditional probability distributions.  Such test spaces are
often called ``semi-classical'' test spaces in the quantum logic
literature \cite{Wilce2000a,Foulis98a}.

(c) If ${\frak A}$ is the collection of all maximal orthonormal
subsets (i.e. orthonormal bases) of a Hilbert space $\H$ of dimension
at least $3$, then $\Omega({\frak A})$ is canonically isomorphic to
the convex set of density operators on $\H$, by Gleason's Theorem.

(d) An interesting model, well known in the
quantum logic literature \cite{Wilce2000a,Foulis98a},
consists of three, three-outcome tests $E = \{a,x,b\}, F =
\{b,y,c\}$ and $G = \{c,z,a\}$, pasted together in a loop. The
extreme points of $\Omega({\frak A})$ are the four dispersion-free
states with supports $\{a,y\}, \{b,z\}$, $\{c,x\}$ and $\{x,y,z\}$,
plus the non-dispersion-free state giving $a,b,c$ all probability
$1/2$ and $x$, $y$ and $z$ probability $0$.

(e) For another example, let ${\frak A}$ consist of the rows and
columns of a $3 \times 3$ array: then $\Omega({\frak A})$ is
the convex set of doubly-stochastic $3 \times 3$
matrices, which is not a simplex -- in spite of the fact that the
pure states, corresponding to permutation matrices, are
deterministic.
 \\

\noindent{\em Observables}\\

By a (discrete) {\em observable} on a system with state-space
$\Omega$, we mean a function $F: x \mapsto F_x$ from a finite set
$E$ into $A(\Omega)$, satisfying (i) $F_x \geq 0$ for all $x \in E$,
and $\sum_{x \in E} F_x = u$. Any state $\omega \in \Omega$ pulls
back along $F$ to a probability weight $p \in \Delta(E)$ via $p(x) = F_x (\omega)$. This provides a dual map $F^*: \Omega \rightarrow \Delta(E)$ defined as $F^*(\omega) = p$.
Note that this definition of an observable generalizes the notion of a Positive Operator Valued Measure (POVM) in quantum theory, rather than the more specialized notion of an observable associated with a self-adjoint operator.

A special case of an observable is a list $(a_1,...,a_k)$ of
positive elements of $A(\Omega)$ that sums to $u$ (in this case, the
mapping $F : \{1,...,k\} \rightarrow [0,u]$ taking $i$ to $a_i$ is
implicit.) Most of the observables considered below will be of this
type.

An observable $F$ is said to be {\em informationally complete}, or
IC, if and only if the set of functionals $\{F_x | x \in E\}$
separates states, i.e., if $F_x (\omega) = F_{x} (\mu)$ for all $x
\in E$ implies $\omega = \mu$ for all states $\omega, \mu \in
\Omega$. (This is equivalent to saying that the dual mapping
$F^{\ast} : \Omega \rightarrow \Delta(E)$ is an affine injection.)
Note that $F$ is IC if and only if $\{F_x | x \in E\}$ spans
$A(\Omega)$. If this set is a {\em basis} for $A(\Omega)$, we shall
say the observable is {\em minimally} IC.
% *ML21/11/06* Rewrote to remove reference.  Can be reinstated later if we find it.
%AW: Added the correct reference, to Singer and Stulpe, J Math Phys 1992.
The following result is not new (see \cite{Singer:1992} for an
infinite-dimensional version), but we include a proof for
completeness.

\begin{Lemma}  Any finite-dimensional state space supports a minimal
informationally complete observable.\end{Lemma}

\noindent{\em Proof:} It suffices to produce a sequence
$(a_1,...,a_n)$ of vectors $a_i \in [0,u]$, with $n =
\dim(A(\Omega))$ distinct entries, summing to the order unit $u$.
Let $B = \{b_1,...,b_n\}$ be any basis for $A(\Omega)$. Without loss
of generality, suppose that $\sum_i b_i = ku$, a multiple of the
order unit. (If not, apply a suitable invertible linear
transformation). Let $c$ be the minimum of $\inf\{b_i(\omega) |
\omega \in \Omega\}$. Then $b_i - cu$ is positive. Now, $\sum_{i}
(b_i - cu) = (k- nc)u$, with $k - nc \geq 0$. Hence, if $a_i = (b_i
- cu)/(k-nc)$, we have $a_i \geq 0$ and $\sum_i a_i = u$. Obviously,
$\{a_i | i = 1,...,n\}$ spans $A(\Omega)$, so $(a_1,...,a_n)$ is a
minimal IC observable. $\Box$\\

\noindent{\em Operations}\\

Any physically performable operation on a system should respect
probabilistic mixtures of states, and hence, should be representable
by an affine mapping $\phi : \Omega \rightarrow \Omega'$, where
$\Omega$ is the state space of the system prior to the operation
being performed, and $\Omega'$ is the post-operation state space.  Generally, the set of allowed operations in a given model could be a strict subset of the set of all affine maps.  This should be familiar from the quantum case given in example (c), since in that case the affine maps are the set of all positive, trace-preserving, linear maps on operators, whereas quantum operations are usually taken to be \emph{completely} positive.  Since we are concerned with proving restrictions on the set of operations available in any model, we assume that all affine maps represent possible operations, but the restrictions obviously still apply to any subset of these maps.

\begin{Lemma}\label{AffLem}  Let $E = (a_1,...,a_n)$ be any observable on
$\Omega$, and let $\delta_1,...,\delta_n \in \Omega'$ be any states
in $\Omega'$. Then the mapping $\phi : \Omega \rightarrow \Omega'$
given by
\[\phi : \omega \mapsto \sum_{i} \omega(a_i)\delta_i\] for all $\omega
\in \Omega$ is affine, i.e., an operation.\end{Lemma}

The proof is routine. Physically, such a process could be
implemented by measuring $E$ and then preparing the indicated state.

Notice that any operation $\kappa : \Omega \rightarrow \Omega'$
determines a dual linear transformation $\kappa^{\ast} : A(\Omega')
\rightarrow A(\Omega)$, given by $\kappa^{\ast}(f)(\omega) =
f(\kappa(\omega))$ for all effects $f \in A(\Omega')$ and all states
$\omega \in \Omega$. This mapping preserves positivity and the order
unit, and hence, allows us to pull observables on $\Omega'$ back to
observables on $\Omega$. (In this connection, notice also that if
$\kappa$ is injective, $\kappa^{\ast}$ will pull informationally
complete observables on $\Omega'$ back to informationally complete
observables on $\Omega$.)\\

\section{Tensor Products}

Given two systems with state spaces $\Omega$ and $\Omega'$, we'd
like to construct a state space to represent a coupled system with
these as components. There is in general no unique way to do this,
but rather there is a spectrum of candidates, bounded by a {\em maximal} and
a {\em minimal} tensor product.\\

\noindent{\bf Definition:} The {\em maximal tensor product} of two
state spaces $\Omega$ and $\Omega'$, which
we'll denote by $\Omega
\otimes \Omega'$, is the set of all
bilinear functionals $\mu :
A(\Omega) \times A(\Omega') \rightarrow {\Bbb R}$ that are (i)
positive on pairs $(a,b)$ with $a, b \geq 0$, and (ii) normalized by
$\mu(u,u') = 1$ (where $u$ and $u'$ are the order-units of
$A(\Omega)$ and $A(\Omega')$, respectively).\\

One can show that the maximal tensor product corresponds to the
largest set of joint probability assignments to measurements on the
two component systems, subject to a ``no-signaling'' condition \cite{Barrett:2005, Foulis:1981, Klay:1987}.

Given states $\alpha \in \Omega$ and $\beta \in \Omega'$, one has a
product state $\alpha \otimes \beta \in \Omega \otimes \Omega'$
given by $(\alpha \otimes \beta)(a,b) = \alpha(a)\beta(b)$ for all
$(a,b) \in A(\Omega) \times A(\Omega')$.\\

\noindent{\bf Definition:} The {\em minimal tensor product} of
$\Omega$ and $\Omega'$ is the the convex hull of the set of product
states in $\Omega \otimes \Omega'$. We term such a convex
combination a {\em separable} state, and accordingly denote the
minimal tensor product by $\Omega \otimes_{sep} \Omega'$. A
non-separable state in $\Omega \otimes \Omega'$ will be termed {\em
entangled}.\\

In the present finite-dimensional setting, $V(\Omega \otimes
\Omega') = V(\Omega) \otimes V(\Omega')$ and $A(\Omega \otimes
\Omega') = A(\Omega) \otimes A(\Omega')$ \cite{Horodecki:2005,
Klay:1987, Ludwig:1983, Ludwig:1985, Wilce:1992}. It follows that
$\Omega \otimes \Omega$ and $\Omega \otimes_{sep} \Omega$ have the
same affine dimension. Hence, every state in $\Omega \otimes \Omega$
can be expressed as an {\em affine} combination $\sum_i t_i \alpha_i
\otimes \beta_i$, where $\sum_i t_i = 1$, but the $t_i$ need not
be positive. \\

 \noindent{\em Examples}\\

(a) If $\Omega$ and $\Omega'$ are both classical state spaces, say $\Omega
= \Delta(E)$ and $\Omega' = \Delta(E')$, then $\Omega \otimes
\Omega' = \Omega \otimes_{sep} \Omega'$, both being isomorphic to
$\Delta(E \times E')$.

(b) If $\Omega$ and $\Omega'$ are the state spaces associated with the
semiclassical binary-input, binary-output test space discussed in
example (b) in section 2, then $\Omega \otimes \Omega'$ supports all
bipartite nonsignaling correlations obtainable with two binary inputs
and two binary outputs.  The extreme points are the local
deterministic states specifying a definite output for each input, and
states supporting nonlocal PR-box type correlations.  On the other
hand $\Omega \otimes_{sep} \Omega'$ only contains local states from
which no Bell-inequality violations can be obtained. More generally,
for any pair of semiclassical test spaces $\Omega \otimes \Omega'$
supports all bipartite nonsignaling correlations with the appropriate
cardinality of inputs and outputs, whereas $\Omega \otimes_{sep}
\Omega'$ contains only local states from which no Bell-inequality
violations can be obtained.

(c) If $\Omega$ and $\Omega'$ are the usual state spaces associated
with complex Hilbert spaces $\H$ and $\H'$, then $\Omega \otimes
\Omega'$ is properly larger than the usual quantum state space
associated with $\H \otimes \H'$
% *ML11/10/06* This probably requires explanation.  I don't think we need to do that here, but I have cited the "Influence free states..." paper, in which I presume it is emplained
 \cite{Foulis:1981,Klay:1987,Klay:1988,Barnum:2005}; the minimal
tensor product, consisting of separable states, is properly smaller.\\

Henceforth, by {\em a} tensor product for state spaces $\Omega$ and
$\Omega'$, we'll simply mean some convex set containing $\Omega
\otimes_{\sep} \Omega'$ and contained in $\Omega \otimes \Omega$.\\

\noindent{\em Remark:} Given affine mappings $\phi : \Omega
\rightarrow \Gamma$ and $\phi' : \Omega' \rightarrow \Gamma'$, there
is a unique affine mapping
\[\phi \otimes \phi' : \Omega \otimes \Omega' \rightarrow \Gamma
\otimes \Gamma'\] satisfying $(\phi \otimes \phi')(\alpha \otimes
\beta) = \phi(\alpha)\otimes \phi'(\beta)$ for all $\alpha \in
\Omega$ and all $\alpha' \in \Omega'$. In particular, there is no
notion of ``complete positivity'' for either the minimal or maximal
tensor product. That is, the tensor product of any two positive
linear mappings remains positive with respect to either the maximal
or the minimal tensor cone.
% *ML11/10/06* I would delete the following sentence, since a generic theory would presumably have a generic tensor product, i.e. not the maximal or minmal tensor product, for which this may not hold.
%(The fact that this {\em isn't} the case
%for the quantum-mechanical tensor product is therefore a {\em
%non}-generic feature of QM.)
\\

\noindent {\em Marginal and Conditional States} \\

A state $\omega \in \Omega \otimes \Omega'$ has well-defined marginal
states $\omega_1 \in \Omega$ and $\omega_2 \in \Omega'$ given, respectively, by
\[a(\omega_1) = (a \otimes u')(\omega) \ \text{and} \ b(\omega_2) = (u \otimes
b)(\omega)\] for all effects $a \in [0,u]$, $b \in [0,u']$. This
fact allows us to define {\em conditional} states $\omega_{2,a}$ and
$\omega_{1,b}$ by
\[\omega_{2,a}(b) := \frac{\omega(a,b)}{\omega_1(a)} \ \ \text{and} \
\ \omega_{1,b}(a) := \frac{\omega(a,b)}{\omega_2(b)}.\] We have the
expected identities
\[\omega(a,b) = \omega_1(a) \omega_{2,a}(b) =
\omega_{1,b}(a) \omega_2(b).\]

The following observation is familiar in the setting of both
classical and quantum probability theory:

\begin{Lemma}
\label{lemma: pure marginal implies product}
If either marginal, $\omega_1$ or $\omega_2$, of a
bipartite state $\omega$ in $\Omega \otimes \Omega'$ is pure (i.e. extremal), then
%\Omega --> \Omega', because later we use cases where we tensor distinct
%state spaces, and nothing else need change (I think!) below...
$\omega = \omega_1 \otimes \omega_2$. \end{Lemma}

\noindent{\bf Proof:} Suppose $\omega_2$ is pure. We wish to show
that $\omega(a,b) = \omega_1(a) \omega_2(b)$ for all effects $a, b
\in [0,u]$. Let $E \subseteq [0,u]$ be any observable. Then we have,
\[\omega_2  = \sum_{a \in E} \omega_1(a) \omega_{2,a}.\] This gives us
$\omega_2$ as a convex combination of the states $\omega_{2,a}$ with
coefficients $\omega_1(a)$. As $\omega_2$ is pure, we have for each
$a \in E$ either $\omega_1(a) = 0$ or  $\omega_{2,a} = \omega_2$; in
either case, $\omega(a,b) = \omega_1(a) \omega_2(b)$ for all  $b \in
[0,u]$. Since $E$ was chosen arbitrarily, this holds also for all $a \in [0,u]$. $\Box$.\\

The tensor product construction can be iterated -- we can form
\[\Omega^{n} := \underbrace{\Omega \otimes \cdots \otimes \Omega}_{n
\ \text{times}}.\] Applying Lemma 3 to this setting, we see that the
``monogamy of entanglement'' \cite{Terhal:2004} is an entirely
generic phenomenon. Thus, for instance, if $\omega$ is a tripartite
state in $\Omega_1 \otimes \Omega_2 \otimes \Omega_3$, then we can
form various marginals, e.g., $\omega_{12} \in \Omega_1 \otimes
\Omega_2$, etc., If $\omega_{12}$ is a pure entangled state, then
$\omega = \omega_{12} \otimes \omega_3$ -- whence, $\omega_{23} =
\omega_2 \otimes \omega_3$ and $\omega_{13} = \omega_1 \otimes
\omega_3$.\\

\noindent{\em Remarks:}

(1) In the context of abstract convex sets, the maximal tensor
product (more usually called the injective tensor product) seems
first to have been discussed by Namioka and Phelps
\cite{Namioka:1969}; see also Wittstock \cite{Wittstock:1974} for a
survey. As a model for coupled physical systems, it was discussed
(implicitly) by Foulis and Randall \cite{Foulis:1981}, Kl\"{a}y,
Randall and Foulis \cite{Klay:1987}, and Kl\"{a}y \cite{Klay:1988}.
(See also \cite{Barnum:2005} and \cite{Wilce:1992}).

% *ML21/11/06* Rewrote this remark in light of the fact that entanglement is defined earlier.
(2) The definition of an entangled state as a state not contained in
$\Omega \otimes_{sep} \Omega'$ naturally generalizes the quantum
definition.  A pure state is entangled iff it has a mixed marginal,
and a mixed state is entangled if it cannot be written as a convex
combination of pure product states.  (See \cite{BKOV2003a,BOSV2005a}
for an even more broadly applicable generalization of this
definition of entanglement to convex operational settings.) With
this definition, it is easy to see from Lemma \ref{lemma: pure
marginal implies product} that any tensor product properly larger
than the minimal one contains entangled states.

\section{ Cloning }

A {\em deterministic cloning procedure} for a state $\alpha \in
\Omega$ involves preparing the system in state $\alpha$, preparing a
second copy of the system in a particular state $\beta$, and
performing an operation on the combined system $\Omega \otimes
\Omega$ that takes the initial state $\alpha \otimes \beta$ the
final state $\alpha \otimes \alpha$. Since the initial ancillary
state $\beta$ is supposed to be fixed, we can equally well regard
such a procedure as an affine mapping $\kappa : \Omega \rightarrow
\Omega \otimes \Omega$ such that $\kappa(\alpha) = \alpha \otimes
\alpha$. One can also consider {\em probabilistic} cloning, in which
there is a non-zero probability that the cloning procedure will
simply {\em fail} (but we will know if it does).
%hb nov 9 06 added ``but we will know if it does''
% *ML11/10/06* I would delete the remainder of the sentence because the reader may get the impression that this is the output if the map fails, and we don't need to know anything about the form of the output for our purposes.
%
% yielding an output of the form $t\alpha \otimes
%\alpha$, where $0 < t < 1$.
Barrett has shown in \cite{Barrett:2005} that universal
probabilistic cloning is generically impossible in
(finite-dimensional) non-classical theories. Here, we consider only
deterministic cloning, and accordingly drop the adjective.

Our aim is to show that a set of states simultaneously cloneable, must
also be sharply distinguishable from one another by a single
observable and vice versa. Our proof of this is essentially just
crystalized folklore: cloning allows us to produce large ensembles of
independent copies of each cloneable state; performing the same
measurement on each of these defines an observable on the original
system, which distinguishes among the cloned states to arbitrary
accuracy, by the law of large numbers.  Conversely, if a set of states
is sharply distinguishable then they may be cloned by measuring the
distinguishing observable and then preparing another copy of the
corresponding state.

%AW: Should the following be a footnote?
This observation has already been made in the quantum case (see
\cite{Chefles:1998} for example) and it has also been noted that the
argument does not seem to depend on the details of quantum
mechanics, which is confirmed by the present result.  However, the
argument need not be true in \emph{all} conceivable frameworks for
physical theories, as it depends on the idea that any state can be
reliably prepared and that distinct states are separated by some
measurement. This is true in the present framework, but theories in
which the notion of state includes ``hidden variables'' provide
counterexamples to this.  As a rather extreme example, consider a
theory just like the ones described here, except that the state of
each system is supplemented by a hidden bit that can have value $0$
or $1$, but which has absolutely no effect on measurement outcomes.
Suppose further that any operation from a single system to a
bipartite composite system copies the value of the hidden bit to
both output systems.  In such a world, we can clone states just as
well as in the present framework, but nevertheless we cannot
distinguish between two states that have differing values of the
hidden bit.

%hb nov 9 06, expanded the following paragraph
In the present framework, the existence of a cloning procedure will
depend not only on the structure of the convex set of states, but
also on what kinds of affine mappings one admits as ``physical''
operations. Indeed, the constant mapping that takes every state in
$\Omega$ to the state $\alpha \otimes \alpha$ is affine; thus, on a
liberal understanding of physical operations, in which any affine
mapping between state spaces is physically realizable, {\em every}
state -- mixed as well as pure -- is (deterministically) cloneable
if we do not demand that the same map clone more than this one
state.\\
% *ML20/11/06* I think it is confusing to start talking about broadcasting here.  This comment would be better in the broadcasting section.  I am omitting it for now.
%Our notion of broadcastability does
%make this liberal allowance for all affine maps, but our strong results on
%sets of states that are {\em not} broadcastable by these means still apply
%when a smaller set of maps is deemed physically reasonable.

\noindent{\bf Definitions:} Call a  finite collection
$\alpha_1,...,\alpha_n$ of states
\begin{itemize}
\item[(a)] {\em  co-cloneable} iff there exists a single cloning map
$\kappa : \Omega \rightarrow \Omega^{2}$ that clones them all, i.e.,
$\kappa(\alpha_i) = \alpha_i \otimes \alpha_i$ for every $i =
1,...,n$, and
\item[(b)] {\em jointly distinguishable} iff there exists an observable $E =
(a_0,....,a_n)$ with $\alpha_i (a_j) = \delta_{ij}$. In this case,
we say that the $\alpha_i$ are {\em distinguishable} by $E$, or that
$E$ is {\em distinguishing} for $\alpha_1,...,\alpha_n$.
\end{itemize}

In discrete classical probability theory, any finite collection of
pure states is jointly distinguishable. It is important to note that,
in general, a pairwise-distinguishable set of states will not be
jointly distinguishable. Indeed, in the case of a binary input, binary
output, semiclassical test space (see Example (b) of section 2), any
two extreme states are distinguishable by one of the two tests, but no
observable will sharply distinguish between any three pure states.
(See also the remark following Corollary \ref{corollary: cloneable
simplex} below.)

In finite-dimensional quantum probability theory, the pure states
corresponding to two vectors $v$ and $w$ are distinguishable in the
foregoing sense iff the vectors $v$ and $w$ are orthogonal. More
generally, we have the following

\begin{Lemma} Quantum states $\rho$ and $\rho'$ are distinguishable iff
the corresponding density operators satisfy $\rho\rho' = \rho' \rho
= 0$.
\end{Lemma}

\noindent{\em Proof:}  $\rho$ and $\rho'$ are distinguishable iff
there exists a self-adjoint operator $0 \leq A \leq \1$ with
$\tr(A\rho) = 1$ and $\tr(A\rho') = 0$. Let  $\rho = \sum_i t_i P_i$
where the $P_i$ are rank- one projections associated with unit
vectors $\v_i$, and where the convex coefficients $t_i$ are all
non-zero. If $\tr(A \rho) = 1$, then, $\sum_i t_i \langle A \v_i,
\v_i \rangle = 1$. Since $\0 \leq A \leq \1$, $0 \leq \langle A
\v_i, \v_i \rangle \leq 1$, so we must have $\langle A \v_i, \v_i
\rangle = 1$ for each $i$. In other words, each $\v_i$ belongs to
the eigenspace of $A$ corresponding to eigenvalue $1$. By the same
argument, if $\rho' = \sum_j r_j Q_j$ is a convex combination of
rank-one projections $Q_j$ (with $r_j > 0$ for all $j$), the vectors
in the range of $Q_j$ must belong to the $0$-eigenspace of $A$.
Accordingly, $P_i \perp Q_j$ for every $i$ and every $j$, so that
$\rho \rho'  = \rho' \rho = 0$. $\Box$. \\

An easy extension of this argument shows that a set of quantum
states is jointly distinguishable iff all pairs $\rho, \rho'$ (with
$\rho \ne \rho'$) of corresponding density operators satisfy
$\rho\rho'=0$. That is, a pairwise distinguishable set of quantum
states is jointly distinguishable. As noted above, this is not
generally the case. This is one of many respects in which quantum
probabilistic models are relatively well-behaved.

\begin{Theorem}  \label{theorem: cloning} In any finite-dimensional
probabilistic theory, using any tensor product, distinct states are
co-cloneable iff they are jointly distinguishable.\end{Theorem}

In outline, the proof is simply the observation that, to distinguish
among the states to any given accuracy, it suffices to produce, by
iterated cloning, a sufficiently large ensemble of independent
copies of each cloneable state, and then to apply to each copy any
observable on which these states have distinct distributions. The
details are as follows:\\

\noindent {\em Proof:} Suppose first that $\alpha_1,...,\alpha_n$
are distinguishable by $E = \{a_1,...,a_n\}$. Define $\kappa :
\Omega \rightarrow \Omega^{2}$ by
\beqa \label{definition of a cloning map}
\kappa(\omega) = \sum_{i=0}^{n} a_i(\omega)\alpha_i \otimes \alpha_i
\eeqa
where $\alpha_0$ is chosen arbitrarily. As observed in Lemma \ref{AffLem}, this
mapping is affine; obviously, $\kappa(\alpha_i) = \alpha_i \otimes
\alpha_i$ for $i = 1,...,n$.

For the converse, we use the fact that---regardless of what tensor
product we use!---cloning maps can be iterated. Let $E \subseteq
[0,u]$ be an informationally complete observable (as afforded by
Lemma 1), and consider the $N$-fold iterated cloning map $\kappa_N :
\Omega \rightarrow \Omega^{2N}$, where $N$ is a large positive
integer. The set $E_{N} := E^{2N}$ is a partition of unity in
$A(\Omega^{2N})$. Every sequence $\x = (x_j,...,x_{2^{N}})$ in
$E_{N}$ determines an empirical distribution $p_{\x}$ on $E$, given
by
\[p_{\x}(x) = \frac{|\{j | x_j = x\}|}{2^{N}}.\]
For each $i = 1,...,n$, let
\[A_{i,N,\epsilon} = \{\ \x \in E_{N} \ | \ \|p_{\x} - \alpha_{i}\| <
\epsilon  \ \},\] where $\| f \|_{E}$ denotes the maximum absolute
value of a function $f$ over $E$. By the weak law of large numbers,
if $\alpha_{i,N} := (\alpha_{i})^{2N} = \kappa_{N}(\alpha)$, then
$\alpha_{i,N}(A_{i,N,\epsilon}) > 1 - \epsilon$ for sufficiently
large $N$.

Let $a_{i,N,\epsilon}$ be the unique functional in $[0,u]$ defined
by $a_{i,N,\epsilon}(\omega) = \kappa^{N}(\omega)(A_{i,N,\epsilon})$
for all $\omega \in \Omega$ (in other words, the pull-back of the
set $A_{i,N,\epsilon}$ along $\kappa^{N}$). We then have
$\alpha_{i}(a_i,N,\epsilon) > 1- \epsilon$ for sufficiently large
$N$. Note that, since only finitely many $\alpha_i$ are involved, we
can choose $N$ large enough to make this hold simultaneously for
{\em all} $i=1,...,N$. We claim that, for sufficiently large $N$ and
sufficiently small $\epsilon$, $\{a_{i,N,\epsilon}\}$ is summable in
$E$, hence, extends to a partition of unity. It is sufficient to
show that $A_{i,N,\epsilon} \cap A_{k,N,\epsilon} = \emptyset$ for
$i \not = k$. To this end, note that since $E$ is informationally
complete, the distinct states $\alpha_{i}$ induce distinct
probability distributions on $E$. In particular, there is some
$\delta > 0$ such that $\|\alpha_i - \alpha_k\|_{E} > \delta$ for
all $i \not = k$. Let $\epsilon < \delta/2$. If $\x \in
A_{i,N,\epsilon} \cap A_{k,N,\epsilon}$, then
\[\|\alpha_i - p_{\x}\|_{E} < \epsilon \ \text{and} \ \|p_{\x} - \alpha_k
\|_{E} < \epsilon,\] so $\|\alpha_i - \alpha_k\|_{E} < 2\epsilon <
\delta$ -- a contradiction. Thus, $A_{i,N,\epsilon} \cap
A_{k,N,\epsilon} = \emptyset$, as claimed.

Now let $a_{0} = \kappa^{\ast}(E_{N} \setminus \bigcup_{i}
A_{i,N,\epsilon})$. We now have an observable $E_{N,\epsilon} =
(a_{i,N,\epsilon} | i = 0,1....,N)$, such that
$\alpha_i(a_{i,N,\epsilon}) > 1-\epsilon$ for each $i$. Since
$[0,e]^{N}$ is compact, we can choose from among the
$E_{N,\epsilon}$ a convergent sequence of observables $E_{m} =
(a_{0,m},...,a_{N,m})$ with $a_{i,m}(\alpha_i) > 1-1/m$ for all $i$.
Thus, for each $i$, the sequence $(a_{i,m})$ converges in $[0,u]$ to
an effect $a_i$ with $a_i(\alpha_i) = 1$. We also have
\[\sum_{i=0}^{N} a_i = \lim_m \sum_{i=1}^{N} a_{i,m} = \lim_m u = u.\] Thus,
$(a_0,...,a_n)$ is a distinguishing observable for
$\alpha_1,...,\alpha_n$, as
advertised. $\Box$.\\

The familiar quantum no-cloning result follows, in view of the
remarks about orthogonality preceding the proof. The following
result shows that only classical systems --  i.e., those the state
spaces of which are simplices -- allow universal deterministic
cloning.

\begin{Cor}
\label{corollary: cloneable simplex}
Suppose that $\alpha_1,..,\alpha_n$ are co-cloneable.
Then the convex hull of $\alpha_1,...,\alpha_n$ in $\Omega$ is a
simplex. Hence, if every finite set of pure (extremal) states in $\Omega$ is co-cloneable then $\Omega$ is a simplex.
\end{Cor}

\noindent{\em Proof:} A simplex is the only finite dimensional
convex set for which each element has a unique decomposition into
extremal states. Hence, let $\alpha_1,...,\alpha_n$ be jointly
distinguishable states, and let $\sum_i s_i \alpha_i = \sum_i t_i
\alpha_i = \omega \in \Omega$, where $s_1,...,s_n$ and $t_1,...,t_n$
are convex coefficients. Let $E = (a_0,a_1,...,a_n)$ be a
discriminating observable for $\alpha_1,...,\alpha_n$. Then $s_i =
a_i(\omega) = t_i$. $\Box$\\

\noindent{\em Remark:} One can certainly construct non-classical
theories in which any {\em pair} of extremal states is
distinguishable, and hence cloneable. For example, consider a
semi-classical test space, that is, a pairwise disjoint collection
of outcome-sets. A pure state on such a test space amounts to a
selection of one outcome per test, and any two such states are
distinguished by any test on which they differ. (Single systems in
both of the theories GNST and GLT considered in \cite{Barrett:2005}
are of this form.)
\\

\section{Broadcasting}

We say that a state $\rho \in \Omega$ is {\em broadcast} by an
affine mapping $B : \Omega \rightarrow \Omega \otimes \Omega$ iff
the bipartite state $B(\rho)$ has marginals equal to $\rho$.
%HB--deemphasized marginals as they're already defined.
% *ML21/11/06* Remarks on cloning are repeated very soon below.  Made an executive decision to move all the discussion there.
%Evidently, cloning is a special case of broadcasting. Indeed, for
%extreme states of $\Omega$, broadcasting reduces to cloning: if
%$\alpha$ is extreme and $B(\alpha)$ has marginals equal to $\alpha$,
%then by Lemma 3, $B(\alpha) = \alpha \otimes \alpha$.
The quantum no-broadcasting result of Barnum et al. \cite{Barnum:1996}
tells us that
two {\em quantum} states are jointly broadcastable iff, regarded as
density operators, they commute. Our aim in this section is to
obtain a characterization of joint broadcastability for arbitrary
systems.

Let $B : \Omega \rightarrow \Omega \otimes \Omega$ be an affine
mapping. We define the marginal mappings $B_1, B_2 : \Omega
\rightarrow \Omega$ by $B_1(\rho)(a) = B(\rho)(a \otimes u)$ and
$B_2(\rho)(b) = B(\rho)(u \otimes b)$.\\

\noindent{\bf Definition:} \label{definition: broadcast} We say that
$\rho \in \Omega$ is {\bf broadcast} by $B$ iff $B_1(\rho) = B_2(\rho)
= \rho$ -- that is, iff $\rho$ is simultaneously a fixed point of both
$B_1$ and $B_2$. Let $\Gamma$ be the set of all states $\rho \in
\Omega$ broadcast by $B$.  Note that $\Gamma$ is a convex
subset of $\Omega$.  Indeed, it is $\Omega$-affine, meaning it is
the intersection of $\Omega$ with an affine subspace.\\

Cloning is a special case of broadcasting. Indeed, for
pure states of $\Omega$, broadcasting reduces to cloning: if
$\alpha$ is extreme and $B(\alpha)$ has marginals equal to $\alpha$,
then by Lemma 3, $B(\alpha) = \alpha \otimes \alpha$.
% *ML21/11/06* Replaced with the more detailed discussion from above.
%Cloning is a special case of broadcasting. For {\em pure} states,
%broadcasting reduces to cloning (since a bipartite state having pure
%marginals is a product state).
Thus, by our no-cloning theorem,
there can be no {\em universally} broadcasting map on a
non-simplicial state space. On the other hand, all states in the
convex hull of a distinguishable set of states can be broadcast,
simply by cloning the extreme points.  To be explicit, let $\rho = \sum_{i}
t_i \alpha_i$ be a convex combination of co-cloneable states
$\alpha_1,...,\alpha_n$, and let
% *ML21/11/06* 1_1 -> a_0.  [] -> () for consistency with earlier notation
$E = (a_0,...,a_n)$ be a distinguishing observable
for $\alpha_1,..,\alpha_n$. Then the very map $\kappa$ used to clone the
$\alpha_i$ in the proof of Theorem \ref{theorem: cloning}, namely,
\[\kappa : \omega \mapsto \sum_i \omega(a_i)\alpha_i \otimes \alpha_i.\]
applied to $\rho$, yields
\[\kappa(\rho) = \sum_i t_i \kappa(\alpha)_i = \sum_i t_i \alpha_i \otimes \alpha_i.\]
Taking the first marginal of this, we have
\[a(\kappa(\rho)_1) = \sum_i t_i a(\alpha_i) = a(\sum_{i} t_i \alpha_i) =
a(\rho);\] similarly, the second marginal is also $\rho$. Thus,
$\kappa$ is broadcasting on
$\Delta(\{\alpha_1,\alpha_2,\ldots,\alpha_n\})$.

In fact, the convexity of the set of states broadcast by any map $B$
shows that any map that broadcasts $\Gamma$'s extreme points
broadcasts $\Gamma$.  If $\Gamma$'s extreme points are extremal in
$\Omega$ then, as mentioned above, a broadcasting map for $\Gamma$
must clone them, but this is not so in general.  Any map of the form
\beqa B: \omega \mapsto \sum_i \omega(a_i) \rho_i\;, \eeqa where
$\rho_i$'s marginals are both equal to $\alpha_i$ and $[a_i]$ as
usual distinguish the $\alpha_i$, broadcasts $\omega \in
\Delta(\{\alpha_i\})$, even though $\rho_i$ may not be $\alpha_i
\otimes \alpha_i$.

% *ML21/11/06* Removed this part of the discussion because the example is not correct.  If it can be fixed then I think it belongs at the end of the paper.

%And even this class of maps is not in
%general the largest class of maps broadcasting $\Delta(\{\alpha_i\})$.
%It is not in general necessary to effectively measure a distinguishing observable,
%``decohering'' the $\alpha_i$, in order to broadcast them.
%For example, a set of commuting
%quantum states with common diagonal basis $\ket{i}$ on a finite
%dimensional Hilbert space $H$, i.e. states
%$\rho = \sum_i \lambda_i \proj{i}$
%with diagonal basis $\ket{i}$ can be broadcast to entangled states
%$\proj{\psi}$, with $\ket{\psi} =
%\sum_i \sqrt{\lambda_i} e^{\sqrt{-1} \phi_i} \ket{i}\ket{i}$, via
%the map $X \mapsto VXV^\dagger$, where the isometry
%$V: H \rightarrow H$ is
%defined by $V: \ket{i} \mapsto e^{\sqrt{-1} \phi_i} \ket{i}\ket{i}$.

If $\Gamma$ is a convex subset of a convex set $\Omega$, then every
affine functional $a \in A(\Omega)$ defines, by restriction, an
affine functional
% *ML11/10/06* a_S->a_\Gamma
$a_\Gamma$ on $\Gamma$. This gives us a natural
positive linear mapping
% *ML11/10/06* a_S->a_\Gamma
$a \mapsto a_\Gamma$ from $A(\Omega)$ to
$A(\Gamma)$, taking the order unit $u \in A(\Omega)$ to the order
unit $u_{\Gamma}$ in $A(\Gamma)$. By a {\em compression} of a convex
set $\Omega$ onto $\Gamma$, we mean an idempotent affine mapping $P
: \Omega \rightarrow \Omega$ having range $\Gamma$. The existence of
a compression implies that the natural mapping $A(\Omega)
\rightarrow A(\Gamma)$ is surjective.

\begin{Lemma} \label{lemma: fixed-point compressions exist} Let $A : \Omega \rightarrow \Omega$
be any affine mapping taking $\Omega$ into itself. Then there exists
a compression of $\Omega$ onto the set of fixed points of $A$.
\end{Lemma}

\noindent{\em Proof:} For each $n \in {\Bbb N}$, let
\[P_{n} = \frac{1}{n} \sum_{k=1}^{n}
A^{k} : \Omega \rightarrow \Omega.\] Since $\Omega$ is compact, we
may assume (passing to a subsequence if necessary) that $(P_{n})$,
converges to a limiting affine map $P : \Omega \rightarrow \Omega$.
If $A(\rho) = \rho$, then clearly $P(\rho) = \rho$; conversely, if
$\rho = P(\mu)$ for some $\mu \in \Omega$,  then we have
\begin{eqnarray*}
A(\rho) & = & \lim_{n \rightarrow \infty} \frac{1}{n} \sum_{k=1}^{n}
A^{k+1}(\mu)\\ & = & \lim_{n \rightarrow \infty} \frac{1}{n}
\sum_{k=1}^{n+1} A^{k} (\mu) - \lim_{n \rightarrow \infty} \frac{1}{n}
A(\mu)\\ & = & \lim_{n \rightarrow \infty} \frac{1}{n} \sum_{k=1}^n
A^k (\mu) - \lim_{n \rightarrow \infty} \frac{1}{n} A(\mu) \\ & & +
\lim_{n \rightarrow \infty} \frac{1}{n} A^{n+1}(\mu) \\
%hb The preceding line has been revised;  see if you think this is the
%correct version
& = & P(\mu) = \rho.
\end{eqnarray*}
Thus, the range of $P$ is exactly the fixed-point set of $A$, as
advertised. Note also that, as $P(\mu)$ is a fixed point of $A$, we
have $P(P(\mu)) = P(\mu)$ for any
$\mu$, i.e., $P$ is idempotent. $\Box$\\

\begin{Lemma} Let $P : \Omega \rightarrow \Omega$ be a compression
of a convex set $\Omega$ onto a convex subset $\Gamma \subseteq
\Omega$. Then (i) $\Gamma \otimes \Gamma$ can be regarded as a
convex subset of $\Omega \otimes \Omega$, and (ii) the mapping $P
\otimes P : \Omega \otimes \Omega \rightarrow \Omega \otimes \Omega$
has range contained in $\Gamma \otimes \Gamma$.
\end{Lemma}

 \noindent{\bf Proof:} We can regard $P$ as a surjective mapping
from $\Omega$ to $\Gamma$. If $a$ is a positive affine functional on
$\Gamma$, then $P^{\ast}(a) := a \circ P$ is an extension of $a$ to
a positive affine functional on $\Omega$. Now for every $\omega$
belonging to $\Gamma \otimes \Gamma$, define a bilinear form
$\overline{\omega} : A(\Omega) \times A(\Omega) \rightarrow {\Bbb
R}$ by $\overline{\omega}(a,b) = \omega(a_\Gamma, b_\Gamma)$; this
is obviously positive and normalized, so $\overline{\omega} \in
\Omega \otimes \Omega$. The mapping $\omega \mapsto
\overline{\omega}$ is clearly affine; it is also injective, by the
aforementioned extension property.  Identifying $\omega$ with
$\overline{\omega}$, we can (and shall) regard $\Gamma \otimes
\Gamma$ as a convex subset of $\Omega \otimes \Omega$. It now
follows (see the remark at the bottom of page 5) that $P \otimes P :
\Omega \otimes \Omega \rightarrow \Gamma \otimes \Gamma$ is a well
defined affine mapping; composing this with the injection $\omega
\mapsto \overline{\omega}$, we have that $P \otimes P$ takes $\Omega
\otimes \Omega$ into itself, with range contained in $\Gamma \otimes
\Gamma$. $\Box$

\begin{Theorem} \label{theorem:broadcasting1}
Let $\Gamma$ be the set of states broadcast by an affine
mapping $B : \Omega \rightarrow \Omega \otimes \Omega$. Then
$\Gamma$ is contained in the simplex generated by a set of
distinguishable states in $\Omega$. \end{Theorem}

\noindent{\em Proof:} Let $\sigma : \Omega \otimes \Omega
\rightarrow \Omega \otimes \Omega$ be the affine isomorphism that
interchanges the two factors. Given the broadcasting map $B : \Omega
\rightarrow \Omega \otimes \Omega$, define another affine mapping
$B' : \Omega \rightarrow \Omega \otimes \Omega$ by $B' = (B + \sigma
\circ B)/2$. Note that $B'$ broadcasts every state $\rho \in
\Gamma$. Call a state $\rho \in \Omega$ {\em symmetrically
broadcastable} iff it is broadcast by $B'$, and denote by $\Gamma'$
the set of all such states. As just observed, $\Gamma \subseteq
\Gamma'$.

Observe that $\rho \in \Gamma'$ iff $\rho$ is a fixed point of the
mapping $B'_{1}$ sending $\rho$ to the marginal $B'(\rho)_{1}$. By
Lemma 5, we have a compression $P$ onto $\Gamma'$.
%hb contraction --> compression
Notice that $P^{\ast} : A(\Gamma') \rightarrow A(\Omega)$ is a
positive linear injection, with $P^{\ast}(u_{\Gamma'}) = u$ (since
$P^{\ast}(u_{\Gamma'})(\omega) = u_{\Gamma'}(P(\omega)) = 1$, since
$P(\omega) \in \Gamma'$.) By Lemma 6, we also have a mapping $Q :
\Gamma' \rightarrow \Gamma' \otimes \Gamma'$ given by
\[ Q(\rho) = (P \otimes P)(B(\rho)).\]
The claim is that this is universally broadcasting on $\Gamma'$. For
if $\rho \in \Gamma'$, we have, for all $a \in [0,u_{\Gamma'}]$,
\begin{eqnarray*}
Q_1 (\rho)(a) & = & Q(\rho)(a \otimes u_\Gamma)\\
& = & ((P \otimes P) B (\rho))(a \otimes u)\\
& = & B(\rho)(P^{\ast} a \otimes P^{\ast} u_\Gamma ) \\
& = & B_1(\rho)(P^{\ast} a) = \rho(P^{\ast} a) \\
& = & P(\rho)(a) = \rho(a) \end{eqnarray*} (using, in the last step,
the fact that $P(\rho) = \rho$, since $\rho \in \Gamma'$). It
follows that $Q_1(\rho) = \rho$; in the same way, one has that
$Q_2(\rho) = \rho$. Since $Q$ is universally broadcasting on
$\Gamma'$, it must in particular broadcast every extreme state
$\alpha \in \Gamma'$. But then Lemma 3 implies that $Q(\alpha)$,
being a state in $\Gamma' \otimes \Gamma'$ with extreme marginals,
must be a product state, namely, $\alpha \otimes \alpha$. Thus, $Q$
is (jointly) {\em cloning} for all of $\Gamma'$'s extreme points. It
follows now from Theorem 1 that these extreme points are
distinguishable in $\Gamma'$ -- hence, also in $\Omega$ (since
any observable on $\Gamma'$ lifts to one on $\Omega$). $\Box$\\

We now have a quantum no-broadcasting theorem as an easy

\begin{Cor}
Let $\Gamma$ be a set of density operators on a Hilbert space $\H$.
Suppose that there exists a positive map $\phi : {\cal B}(\H)
\rightarrow {\cal B}(\H)$ broadcasting each $\rho \in \Gamma$. Then
the operators in $\Gamma$ are pairwise commuting.\end{Cor}

\noindent{\em Proof:} By Theorem 2, $\Gamma$ is contained in a
simplex generated by distinguishable -- hence, by
Lemma 4, commuting -- density operators. It follows that the operators in $\Gamma$
also commute. $\Box$\\

\noindent{\em Remarks:}

(1) The standard {\em quantum} no-broadcasting
theorem applies to a {\em completely positive} broadcasting map. Our
result gives, in the form of the above Corollary, a stronger formulation:
that no {\em positive} map between matrix algebras can broadcast two
non-commuting states.

(2) As stated, Theorem 2 tells us little about the convex structure
of the set $\Gamma$ of states broadcast by a map $B$ (since any
convex set can be embedded in a simplex). Combining it with the
simple observation made above near Definition \ref{definition:
broadcast} that $\Gamma$ is $\Omega$-affine, we can say more:  that
$\Gamma$ is an affine section of a simplex generated by
distinguishable states. Our next result is that $\Gamma$ in fact
{\em is} a simplex generated by distinguishable states.

% *ML21/11/06* Added "jointly" in statement of the theorem
\begin{Theorem} \label{theorem:HowardsThe} Let $\Gamma$ be the set of states broadcast by an affine
mapping $B : \Omega \rightarrow \Omega \otimes \Omega$. Then
$\Gamma$ is a simplex generated by jointly distinguishable
states in $\Omega$. \end{Theorem}

% *ML21/11/06* Rewrote the proof to make the distinction between the states/maps in \Omega and the vectors/stochastic matrices clearer
\noindent{\em Proof:} We maintain the definitions used in the proof of
Theorem \ref{theorem:broadcasting1}.  Any state $\omega \in \Gamma'$
has a unique representation $\omega = \sum_i \omega_i \alpha_i$ as a
convex combination of the extremal points $\alpha_i$ of the simplex
$\Gamma'$.  Let
% *ML20/11/06* i starts from 0 rather than 1 in Alex's earlier definition of joint distinguishability.  It's used in the no-cloning proof and I've maintained it here for consistency.
$[a'_i]_{i=0}^{n}$ be a measurement that distinguishes
the vertices of $\Gamma'$.  The $a'_0$ outcome has probability $0$ on all states in $\Gamma'$, so we may set $a_1 = a'_0 + a'_1$ and $a_i = a'_i$ for $2 \leq i \leq n$ to obtain an observable $[a_i]_{i=1}^n$ that still satisfies $\alpha_i(a_j) = \delta_{ij}$.   This observable can be used to define a restriction map $r: \Omega \rightarrow \Gamma$ via
\begin{equation}
r(\omega) = \sum_{i=1}^n \omega(a_i) \alpha_i,
\end{equation}
which is affine and surjective.
For any $\omega \in \Omega$, this induces a unique
``reduced state'' $\omega^r \in \Gamma'$ defined as $\omega^r = r(\omega)$.  All these ``reduced states'' $\omega^r$ are determined uniquely by an $n$-vector $v^\omega$ of probabilities, with components $v^\omega_i = \omega(a_i)$.

Any state $\omega \in \Gamma$ satisfies $(B_m(\omega))^r =
B_m(\omega) = \omega$ for $m = 1,2$.  Therefore $B_m(\omega) =
(B_m(\omega))^r = (\sum_i \omega_i B_m(\alpha_i))^r = \sum_i
\omega_i (B_m(\alpha_i))^r$.  Since $(B_m(\alpha_i))^r \in \Gamma'$,
the restriction to $\Gamma'$ of the map $\omega \mapsto
(B_m(\omega))^r$ is a classical stochastic map on the simplex
$\Gamma'$.  This map can be represented as a column stochastic
matrix $M_m$ that acts on the vector $v^\omega$.  The $i$th column
of $M_m$ is just the vector representative of the image of the
vertex $\alpha_i$ under the map $B_m$, i.e. $v^{B_m(\alpha_i)}$.
Thus a state $\omega \in \Gamma'$ is broadcastable if and only if
$M_m v^\omega = v^\omega$ for $m = 1,2$, that is, if $v^\omega$ is
in the intersection of the fixed-point subspaces of both stochastic
matrices $M_m$.  We can understand these fixed point spaces using
the extension of the Perron-Frobenius theory of eigenvectors and
eigenvalues of irreducible nonnegative square matrices to the case
of general (i.e. possibly reducible) nonnegative square matrices.
Appendix \ref{appendix:broadcasting} summarizes this theory and
proves two Lemmas we use.  Lemma \ref{lemma: fixed points of
stochastic maps}, following easily from the extended
Perron-Frobenius theory, gives a basis for the space of fixed points
of a stochastic map consisting of disjointly supported nonnegative
vectors, which correspond to distinguishable states when normalized.
The main technical work of the present proof is in deriving from
this Lemma \ref{lemma: intersection of fixed point spaces of
stochastic maps}, stating that the intersection of the fixed-point
spaces of two such stochastic matrices also has (when it is not
$\{0\}$) a basis of disjointly supported nonnegative vectors $v^I$,
so that the set of normalized states that are fixed points of both
maps is the simplex $\Delta(\{v^I\})$ generated by these
distinguishable states.    Since we established above that $\Gamma$
is the set of states fixed by two stochastic maps, it is a simplex
generated by distinguishable states.  (If the intersection is
$\{0\}$ (as it will be for a generic map $B$), the $\Gamma =
\Delta(\emptyset)=\emptyset$, which we view as a degenerate case of
a simplex generated by a set of distinguishable states.)  \QED \\

\noindent{\em Remark:} Although for a given $B$ both $\Gamma'$ and
$\Gamma$ are simplices generated by distinguishable states, it is
easily shown by example that $\Gamma$ may be a proper subset of
$\Gamma'$. For instance, let $\Omega = \Delta(\{\alpha_1,
\alpha_2\})$ and let $B: \alpha_1 \mapsto \alpha_1 \otimes \alpha_2,
\alpha_2 \mapsto \alpha_2 \otimes \alpha_1$. Then $\Gamma =
\emptyset$ while $\Gamma' = \{(\alpha_1 + \alpha_2)/2\}$.

% *ML21/11/06* Removed remark (2) because it refers to the earlier example that I removed.  A paragraph making a similar point may be added to the conclusion making similar comments if deemed necessary.
%(2) The process of broadcasting results in the extremal
%marginal states $\omega \in \Gamma$ on
%each subsystem being decohered from each other in the final states, in the sense that
%the information about which of these states we have in system $A$ exists in system $B$.  As the example preceding Lemma \ref{lemma: fixed-point compressions exist}
%of coherent quantum broadcasting of a basis illustrates, though,
%the broadcast states $B(\alpha_i)$ themselves (where $\alpha_i$ are the vertices of
%$\Gamma$) need not be decohered by the broadcasting process even though there is always
%a broadcasting map that decoheres them.  It would be interesting to formalize the notion
%of decoherence in general theories
%so the question of when a coherent broadcasting procedure exists could
%be investigated in general in the present framework.

\section{Conclusions}

In order to understand the nature of information processing in quantum
mechanics, it is important to be able to delineate clearly those
probabilistic and information-theoretic phenomena that are indeed {\em
essentially} quantum, from those that are more generically
non-classical.
% *ML21/11/06* I'm just removing the offending sentence for now.  Alex can reinstate with references if he wants to.
%It has been recognized for a long time in the
%quantum-logical literature that the general phenomenon of {\em
%entanglement} is of the latter kind.  [HB--we need references here.
%How is entanglement defined in this literature?]
We have established here that several specific features of quantum
information are generic: entanglement monogamy, and, in
finite-dimensional theories, the connection between cloning and
state-discrimination and the no-broadcasting theorem.

One might wonder at this point whether {\em every} qualitative
result of
% *ML11/10/06* Again, I have removed the reference to entanglement.
quantum information will turn out to be similarly generic, either in non-classical
theories or in all theories.
%HB--I put in ``or in all theories'' because of the worry that teleportation IS
%possible classically.  If there are different features of teleportation involved
%than I have in mind, maybe we should reinstate things the old way, but
This is not the case, however. For example, not every
finite-dimensional probabilistic theory allows for teleportation
(this is shown in \cite{Barrett:2005} and also follows from the
results of \cite{Short:2005a} on entanglement swapping.)

Finally, it is worth commenting on the program of deriving quantum
theory from information theoretic axioms \cite{Fuchs:2002, Fuchs:2003,
Clifton:2003} in the light of the present work.  Any such attempt must
begin with a framework that delineates the set of theories under
consideration.  The framework must be narrow enough to allow the
axioms to be succinctly expressed mathematically, but broad enough
that the main substantive assumptions are contained in the axioms
rather than in the framework itself.  The generalized probability
models discussed in this paper would appear to be a natural choice for
this task.

In \cite{Clifton:2003}, Clifton, Bub and Halvorson attempt an
information theoretic axiomatization within a $C^*$-algebraic
framework, which is narrower than the framework adopted here.  In
fact, the $C^*$ framework is already very close to quantum theory,
in the sense that all theories in the framework have Hilbert space
representations.  In the finite dimensional case, quantum theory,
classical probability and quantum theory with superselection rules
are the only options available.%%AWK - fix
The information theoretic axioms used in \cite{Clifton:2003} are:
no-signaling, no-broadcasting and no-bit-commitment. From these it
is shown that there must be noncommuting observables in the theory
and there must be some entangled states.  Given the restricted
nature of the $C^*$ framework, this already yields a theory that
looks quite close to quantum theory.

In contrast, the generalized probabilistic framework adopted here
automatically satisfies no-signaling, and we have shown that
no-broadcasting is generically true of any nonclassical model.  Such
generic models can look very different from quantum theory.  For
example, they include models that support super-quantum
correlations. An open question is whether no-bit-commitment is also
generic in the present framework, and it is possible that it does
place nontrivial constraints on the choice of tensor product.
Nevertheless, it seems unlikely that these three axioms alone would
get one particularly close to quantum theory.
% *ML20/11/06* Rewrote this sentence.
In the light of this, it seems that the best
hope for future progress in axiomatization would be to supplement or
replace these axioms with things
that do not appear to be generic, such as the existence of a
teleportation protocol.\\

% *ML21/11/06* Modified acknowledgments to include yet more funding statements required by Cambridge.
\noindent{\bf Acknowledgments:} Research at Perimeter Institute is
supported in part by the Government of Canada through NSERC and by the
Province of Ontario through MEDT.   At Cambridge, ML was supported by the European
Commission through QAP, QAP IST-3-015848, and through the FP6-FET Integrated Project SCALA, CT-015714.

\bibliographystyle{plain}
\bibliography{GenClone}

\appendix

\section{Perron-Frobenius Theory, Fixed Points of Classical Stochastic Maps,
and Lemmas Used in Proving Theorem
\ref{theorem:HowardsThe}}

\label{appendix:broadcasting}

By a nonnegative matrix (or row or column vector) we mean
one with real nonnegative entries.  By a semipositive matrix or vector, we mean one
with nonnegative entries at least one of which is positive, and by a positive matrix
or vector, we mean one for which every entry is strictly positive.
A nonnegative matrix is called
{\em reducible} if there exists a permutation matrix $P$ such that
$P M P^t$ has the form:
% *ML21/11/06* Changed A's to M's for consistency with rest of discussion
\beq
\left(
\begin{array}{lll}
M^{11} & 0 \\
M^{12} & M^{22}
\end{array}
\right) \; ,
\eeq
{\em irreducible} if there does not.
Some such permutation $P$ will put a general nonnegative square matrix $M$ in
{\em Frobenius normal form}

\beq
\left(
\begin{array}{lllll}
M^{11} & 0 & 0 & \cdots & 0 \\
M^{21} & M^{22} & 0 & \cdots & 0 \\
M^{31} & M^{32} & M^{33} & \cdots & 0 \\
\vdots & \vdots & \vdots & \cdots & 0 \\
M^{K1} & M^{K2} & M^{K3} & \cdots & M^{KK}
\end{array}
\right)
\eeq
% *ML21/11/06* Changed subscript to superscript.
where each diagonal block $M^{II}, I \in \{1,...,K\}$ is irreducible.

The standard Perron-Frobenius theory applies to irreducible nonnegative
square matrices
$M$, guaranteeing a strictly positive eigenvector with
a real positive eigenvalue $\rho(M)$ greater than or equal to the modulus
of any other eigenvalue, real or complex (thus $\rho(M)$ is the spectral radius of $M$).

A result explicitly stated and proved in \cite{Cooper:1973}, and also stated in
\cite{Schneider:1986} (where its proof is said to be essentially present in
Frobenius \cite{Frobenius:1912}) partially
characterizes the real nonnegative eigenvectors of general (possibly reducible) nonnegative
square matrices that correspond to positive eigenvalues.  The eigenvalues of such
nonnegative eigenvectors are $\rho_I := \rho(M^{II})$, and for each diagonal
block $M^{II}$ in the Frobenius normal form of $M$
having a given $\rho_I$, there is an eigenvector $v^I$ whose components
with indices (after the permutation that gives Frobenius normal form)
in block $I$ and above are nonnegative,
and whose lower-indexed components are zero.
% *ML21/11/06* I found the use of ''has access to'' before its definition confusing, so I have reordered this section a bit.
It is also possible to characterize the eigenvectors $v^I$ in a way
which is independent of Frobenius normal form by introducing the following terminology.
An index $i$ {\em has access to} an index $j$ if
there is some finite power $p$ such that $(M^p)_{ij} > 0$.
In the context of
column-stochastic matrices interpreted as transition matrices, this means that
probability can eventually leak from state $j$ to state $i$ (note the directionality,
which is not obvious from the term ``has access to'').  Equivalently, $i$ has access to $j$ if
in the directed ``transition graph'' having edges $(i,j)$ (thought of as
directed ``from $i$ to $j$'') where, and only where,
$M_{ij} \ne 0$ (note again the nonintuitive directionality opposite the flow of probability),
there is a (directed) path from $i$ to $j$.
The indices in a given subset $I$, on which an eigenvector $v^I$ has positive components, can be characterized as mutually having
access to each other (a condition which identifies those subsets without the need to
mention Frobenius normal form as we did above).
Finally, the eigenvectors with a given real positive
eigenvalue $\lambda$ are precisely the real semipositive linear
combinations of the eigenvectors, among those whose existence is asserted above,
having eigenvalue $\lambda$.

% *ML21/11/06* Reworded and changed x to v for consistency of notation.
The next result concerns
the fixed point states,
that is to say the real nonnegative normalized eigenvectors $v$ ($\sum_i v_i = 1$) with
eigenvalue-$1$ of the
column-stochastic matrix $M$.

\begin{Lemma}\label{lemma: fixed points of stochastic maps}
A column-stochastic matrix $M$ may be put into Frobenius normal form
in such a way each of its fixed point states is supported precisely on one
of the $L \le K$ blocks numbered $K-L+1,...K$.  The restriction of $M$
to these blocks will then be block-diagonal.
\end{Lemma}

\noindent{\em Proof:}
Without loss of generality suppose $M$ is in Frobenius normal
form,
with blocks $M^{IJ}, (I,J \in \{1,...,K\}$.

The real positive eigenvalues of a column-stochastic matrix must be
equal to $1$ (because it preserves normalization).  Those of an
irreducible properly substochastic matrix (i.e. one for which all
column sums are less than $1$, and at least one strictly so) must be
strictly less than $1$.

$M_{KK}$ is
column-stochastic, so it follows easily from the irreducible Perron-Frobenius
theory that $\rho_{K}=1$ and there is an eigenvector whose support
is $K$ with eigenvalue $1$.  For any other diagonal block $M^{LL}$ to have
an eigenvalue-$1$ nonnegative eigenvector, it must be the case that all blocks $M^{LM}$
below it ($M < L$) are zero matrices, for if one of them is not, then $M^{LL}$
is properly column-substochastic.  Any such diagonal blocks $M^{LL}$ with
$\rho(M^{LL})=1$ can be put at the end of the ordering of blocks (indeed, in arbitrary
order at the end) by an index permutation preserving Frobenius normal form.  Assume
this has been done, and let
them be blocks $K-L+1$ through $K$.  Thus by the Cooper/Frobenius result discussed above
Lemma \ref{lemma: fixed points of stochastic maps} $M$ has $L$ disjointly supported
fixed-point eigenvectors, one supported on each of the subsets $K-L+1, ..., K$.  The
indices belonging to $1,...K-L$ thus correspond to vertices on which the fixed
points of $M$ have zero support.
\QED

\begin{Lemma}\label{lemma: intersection of fixed point spaces of stochastic maps}
Let $M_1$, $M_2$ be two column-stochastic matrices.  The intersection
of their fixed-point subspaces is spanned by a set of distinguishable
states, so the set of normalized states that are fixed points of both
maps is a simplex generated by distinguishable states.
\end{Lemma}

\noindent{\em Proof:}

We cannot necessarily put both $M_1$ and $M_2$ in Frobenius normal
form simultaneously.  However, the block indices in the Frobenius
normal form of $M_m$ correspond (for each fixed $m$) to a partition of
the set of vertex indices into subsets.

Thus each map's fixed-point space is defined by a partition $\Lambda_m$ of
the vertices into a set $Z_m$ on which it has no support, and sets,
for which we use variables $I,I',...$ for $m=1$, and $J,J',...$ for
$m=2$, of vertices each of which supports a strictly positive
fixed-point vector $v^I$ (resp. $w^J$), with components $v^I_k$
(resp. $w^J_k$).  Thus e.g. $v^I_k = 0$ whenever $k \notin I$.  We
will also define vectors $v = \sum_I v^I$ with components $v_k$, ($w =
\sum_J w^J$ with components $w_k$).  We will use the notation $I(l)$
to mean the subset of the pertinent partition to which the
vertex-index $l$ belongs.

If $\omega$ is in the intersection of the fixed-point spaces of $M_{1,2}$ then
there exist nonnegative $\lambda_I$, $\mu_J$ such that
\beqa \label{constraints on the state}
\omega = \sum_I \lambda_I v^I = \sum_J \mu_J w^J\;.
\eeqa
The first way of expressing $\omega$ enforces that it is a fixed point of $M_1$,
the second, that it is a fixed point of $M_2$.
We now give a procedure for expressing the condition $\omega =
\sum_J \mu_J w^J$ as further
constraints on the $\lambda_I$'s taking the form that for some of the
$I$, $\lambda_I$ must be zero, while some of the ratios $\lambda_I / \lambda_{I'}$
are fixed by the data
% *ML21/11/06* Changed w^J_K to w^J.  I think this was a typo.
$\mu_J, w^J$ when $I,I'$ are both incident on the same $J$.

To do this, it will be useful to define some relations $R,S$ on
$\Lambda := \Lambda_1 \union \Lambda_2$.  We say $G ~R~ H ~{\rm iff}~ G \intersect H \ne 0$.  $R$ is reflexive and symmetric.  Let $S$ be its
transitive closure (i.e. $G ~S~ P$ iff there is a finite chain
$H_1,...H_n$ such that $G ~R~ H_1 ~R~ H_2 ~R~ \cdots ~R~ H_n ~R~ H$).
$S$ is an equivalence relation, so its equivalence classes $[I]_S, [J]_S$
partition
$\Lambda$.  Moreover, it is easy to see that its restrictions $S_1,
S_2$ to $\Lambda_1, \Lambda_2$ are also equivalence relations, and for
any given equivalence class $[I]_S$ or $[J]_S$ of $S$, the equivalence
classes $[I]_{S_1}$ or $[J]_{S_2}$ satisfy $\union \union [I]_{S_1} = \union \union
[I]_{S}$, (or $\union \union [J]_{S_2} = \union \union [J]_S$), i.e. the sets in
them contain the same vertices.

A fact that will be useful below is that if $\lambda^I=0$ (or $\mu^J=0$), then
$\lambda^{I'} = \mu^{J'} =0$ for all $I', J' \in [I]_S$ (or $[J]_S$).  The
reason is that $\lambda_I = 0$ implies $\omega_k = 0$ for all $k \in I$,
so for such $k$, $\mu_J w^J_k = 0$, implying (since $w^J_k >0$) $\mu^J=0$.
In other words, $\lambda_I = 0$ and $I ~R~ J$ imply $\mu_J=0$; the
same argument shows that $\mu_J=0$ and $J ~R~ I'$ implies $\lambda_{I'}=0$; thus
the same statements hold with $S$ in place of $R$ and we see that
zero coefficients for $I$ (or $J$) propagate throughout $[I]_S$ (or $[J]_S$).

Note that if $I \intersect Z_2 \ne 0$ then $\lambda^I = 0, \mu^I=0$
for $I, J \in [I]_S$.  This
is because the vectors $v^I$ have positive components $v^I_k$ for $k
\in I$, but for $k \in Z_2$ we have $\omega_k = 0$, which therefore
requires $\lambda^I=0$; the above observation then applies.

Let $Z'$ be $Z_1$ plus the set of all the vertices that this argument shows to have
$\omega_k = 0$, and $\Lambda_m'$ the partitions of the remainder of
the vertices agreeing with $\Lambda_m$.

Recall from (\ref{constraints on the state}) that the components of $\omega$ satisfy:
\beqa
\mu_{J(k)} w_k = \lambda_{I(k)} v_k\;,
\eeqa
for all $k$.  Thus if $I \intersect J \ne 0$
then either $\mu_J, \lambda_I = 0$ or
$v_k/w_k$ for $k \in I \cap J$ is some constant $\beta_{IJ} :=
\mu_{J}/\lambda_{I}$ independent of $k$.
So $\lambda_I$ must be zero if there is any $J$ with $J \cap I \ne \emptyset$
for which $v$ is not proportional to $w$ on $J \cap I$.  As before,
the upshot is that if an equivalence class $X$ of
$S$ contains sets $J,I$ such that $v$ is not proportional to $w$ on $I \intersect J$,
the coefficients of all sets in $X$ must be zero. This constraint removes
more indices from the subset on which joint fixed-points can be supported
(implying the fixed-points lie in the subsimplex of
$\Gamma'$ with those vertices deleted).  We therefore define
$Z'', \Lambda_m''$ similarly to $Z', \Lambda_m'$.
Now let
$J(k) = J(l)= J$ but $I(k) = I \ne I(l) = I'$ and consider
\beqa
\frac{v_k/w_k}{v_l/w_l} = \frac{\mu_{J(k)}}{\lambda_{I(k)}} \frac{\lambda_{I(l)}}
{\mu_{J(l)}}\;.
\eeqa
If $\lambda_I \ne 0$ then $\mu_J, \lambda_{I'} \ne 0$ and we get the
requirement:
\beq \label{the requirement}
\frac{\lambda_{I(l)}}{\lambda_{I(k)}} = \frac{v_k}{w_k} \frac{w_l}{v_l}\;,
\eeq
i.e.
\beq
\frac{\lambda_{I'}}{\lambda_{I}} = \beta_{IJ}/\beta_{I'J}\;,
\eeq
As promised, some of the constraints coming from $M_2$ have fixed the ratio
of $\lambda_I$ and $\lambda_{I'}$.  Any $J' \ne J$ incident on both $I$
and $I'$ must give rise to the same ratio $\lambda_I'/\lambda_I$; that is,
\beq
\beta_{IJ}/\beta_{I'J} = \beta_{IJ'}/\beta_{I'J'}\;.
\eeq
Should this not be the case, our assumption that $\lambda_I \ne 0$ must be
false, so all $\lambda_{I''}=0$ for $I'' \in [I]_{S_1}$.

Thus, the ratios $\lambda_{I'}/\lambda_{I}$ are fixed
to $\gamma_{I'I} := \beta_{IJ}/\beta_{I'J}$ within those $S$-equivalence class
for which the RHS is independent of $J$, while all $\lambda_I=0$
in the other $S$-equivalence classes.  No constraints on the $\lambda_I$
arise across $S$-equivalence classes.  We add the zeroed-out vertices to
$Z''$ to obtain $Z'''$, and similarly obtain $\Lambda_1'''$ as the
remaining $S_1$-equivalence classes.

Some obvious consistency conditions must be satisfied by
the ratios $\gamma_{I I'}
= \lambda_I/\lambda_{I'}$ thus obtained, namely the transitivity conditions:
\beq
\gamma_{II'} \gamma_{I' I''} = \gamma_{I I''}\;.
\eeq
It may be the case that one side of this is defined while the other side is
not, because, for example, although some $J$ is incident on both $I$ and $I'$,
no $J$ is incident on both $I$ and $I''$, in which case no further constraint
arises;  but when all are defined, we have (recalling the definition of $\beta_{IJ}$)
that:
\beq
\frac{v_k/w_l}{v_{k'}/w_l}  \frac{v_{k'}/w_{l'}}{v_{k''}/w_{l'}}
=   \frac{v_k/w_l}{v_{k''}/w_l} \;.
\eeq
Canceling, we obtain an identity so no further constraints arise.

We have just expressed all the constraints arising from $\omega =
\sum_J \mu_j w^J$ as constraints on the $\lambda_I$.
The other constraint $\omega = \sum_I \lambda_I v^I$ gives $\omega$ as a convex
combination of distinguishable states $v^I$, i.e $\omega \in \Delta(\{v^I\})$.
It is evident
that fixing $\omega_k=0$ for $k \in Z'''$ just says the states are in
a subsimplex of $\Delta(\{v^I\})$, while fixing the ratios of vertices $v^I$
within the elements of a partition just says that the states are convex
combinations of a particular set of disjointly supported, and
therefore still distinguishable, states in this subsimplex.

%%  At this point it is clear that we have shown $\Gamma$ to be
%% a simplex of distinguishable
%% states.  Throwing out the vertices in $Z$ gave us a subsimplex of $\Gamma'$, and
%% the constraint
%% $\omega = \sum_I \lambda_I v^I$ expresses $\omega$ as an arbitrary convex combination
%% of states $v_I$ which are distinguishable because supported on disjoint subsets of the
%% distinguishable vertices of $\Gamma'$; that is, $\omega$ belongs to a simplex $\Delta$ of
%% distinguishable states in $\Gamma'$.  The remaining constraints from $M_2$ we have just
%% shown result in setting some of the $\lambda_I$ to zero and fixing their ratios on some
%% disjoint subsets of the vertices of $\Delta$, which still results in a simplex of
%% distinguishable states in $\Gamma'$.
%%  $\Box$

%\noindent {\em Remark:}

To be rigorous we give an explicit expression for
$\omega$ as a convex combination of distinguishable
states.  Without loss of generality suppose that $\sum_k v^I_k =
\sum_k w^J_k = \sum_I \lambda_I = \sum_J \mu_J = 1$, so that
$\omega$, $v^I$, and $w^J$ are normalized states.
Picking representatives $\hat{I} \in [\hat{I}]$
from each element $[\hat{I}]$ of the partition
$\Lambda_1'''$ we begin with
$\omega = \sum_{I \in \union \Lambda_1'''} \lambda_{I} v^I$ and impose the constraints,
getting:
\beqa \label{expanding omega}
\omega =  \sum_{[\hat{I}]} \lambda_{\hat{I}}
\sum_{I' \in [\hat{I}] } (\lambda_I / \lambda_{\hat{I}}) v^I \equiv
\sum_{[\hat{I}]} \lambda_{\hat{I}} \sum_{I' \in [\hat{I}] }
 \gamma_{I \hat{I}} v^I\;.
\eeqa
Define normalized vectors
\beqa \label{normalized vectors}
v^{[\hat{I}]} := ( \sum_{I' \in [\hat{I}]}
 \gamma_{I \hat{I}} v^I) / ( \sum_{I' \in [\hat{I}]}
 \gamma_{I \hat{I}})\;,
\eeqa
and scalars
\beqa \label{normalized scalars}
\lambda'_{[\hat{I}]}
:= \lambda_{\hat{I}}  ( \sum_{I' \in [\hat{I}]}
 \gamma_{I \hat{I}})\;.
\eeqa
To see that these definitions are independent of the
choice of representative $\hat{I}$ of $[\hat{I}]$,
recall (cf. (\ref{the requirement})) that
\beq
\gamma_{I\hat{I}} :=
\frac{\lambda_I}{\lambda_{\hat{I}}} = \frac{v_p}{w_p} \frac{w_l}{v_l}\;,
\eeq
for any $p \in \hat{I}$, $l \in {I}$ (independently of our choice of
such $p,l$).
Now from (\ref{normalized vectors}),
\beqa
v^{[\hat{I}]}_k =
\frac{\gamma_{I\hat{I}} v^I_k}{\sum_{I' \in [\hat{I}]} \sum_{k \in I'}
\gamma_{I'\hat{I}} v^{I'}_k} \equiv
\frac{\gamma_{I\hat{I}} v^I_k}{\sum_{I' \in [\hat{I}]}
\gamma_{I'\hat{I}}} \;,
\eeqa
and we see that the $\hat{I}$ dependence, which is only
through the factor $\gamma_{I\hat{I}}$ on top and
$\gamma_{I'\hat{I}}$ on the bottom,
takes the form of factors $v_p/w_p$ for some $p \in \hat{I}$ on
the top and bottom, which cancel
establishing the claimed independence from the choice of
$\hat{I} \in [\hat{I}]$.  Also,
\beqa
\lambda'_{[\hat{I}]} := \lambda_{\hat{I}} \sum_{I \in [\hat{I}]}
\gamma_{I\hat{I}} \nonumber \\
= \lambda_{[\hat{I}]} \sum_{I \in [\hat{I}]} (\lambda_I/\lambda_{\hat{I}})
= \sum_{I \in [\hat{I}]} \lambda_I\;,
\eeqa
showing that this too depends only on $[\hat{I}]$.

With these definitions, (\ref{expanding omega}) becomes:
\beqa
\omega  = \sum_{[\hat{I}]} \lambda_{[\hat{I}]}' v^{[\hat{I}]}\;.
\eeqa

Since the sets of vertices
$\cup \cup {[\hat{I}]}$ supporting each $v^{{[\hat{I}]}}$ are disjoint,
the $v^{{[\hat{I}]}}$ are distinguishable, and since in addition the
nonnegative coefficients
$\lambda_{{[\hat{I}]}}$ are free except for overall normalization, $\Gamma$ is the
simplex $\Delta(\{v^{[\hat{I}]}\}_{[\hat{I}]})$
with distinguishable vertices $v^{{[\hat{I}]}}$.
\QED

\end{document}